\magnification=\magstep1
\baselineskip 0.6truecm
\parskip=0.2truecm
\parindent=.5truecm
\raggedbottom
 
\newcount\fcount \fcount=0

\def\ref#1{\global\advance\fcount by 1 \global\xdef#1{\relax\the\fcount}}
\def\pp{\parshape 2 0truecm 16.5truecm .5truecm 14.5truecm}

\def\book #1;#2;#3{\par\pp #1, {\it #2}, #3}
\def\rep #1;#2;#3{\par\pp #1, #2, #3}

\def\simlt{\lower.5ex\hbox{$\; \buildrel < \over \sim \;$}}
\def\simgt{\lower.5ex\hbox{$\; \buildrel > \over \sim \;$}}

\def\np{\vfill\eject}
\def\cl{\centerline}
\def\noi{\noindent}
\def\bs{\bigskip}
\def\lsim{\hbox{\raise 1 truemm \rlap{$<$}\lower 1truemm \hbox{$\sim$}\ }}
\def\lsim{\hbox{\raise 1 truemm \rlap{$>$}\lower 1truemm \hbox{$\sim$}\ }}

\cl {\bf CMB RADIATION POWER SPECTRUM IN CDM OPEN UNIVERSES} 
\cl {\bf UP TO 2nd ORDER PERTURBATIONS}
\bs
\bs

\centerline {Jose Luis Sanz, Enrique Mart\'\i nez-Gonz\'alez} 
\bigskip
 
\cl {Instituto de F\'\i sica de Cantabria,} 
\cl {Consejo Superior de Investigaciones Cientificas-Universidad de Cantabria, 
Santander, Spain}
\bigskip
 
\cl{ Laura Cay\'on, Joseph Silk}
\cl{ Departments of Astronomy and Physics, and}
\cl{Center for Particle Astrophysics}
\cl{University of California, Berkeley, California 94720}
\bigskip 
\cl{Naoshi Sugiyama}
\cl{Department of Physics and Reseach Center for the Early Universe,
Univesity of Tokyo, Tokyo 113, Japan}
\bs
\bs

\cl{\it ABSTRACT}
\bs
\bs
A second--order perturbation theory approach is developed to 
calculate temperature anisotropies in the cosmic microwave background.
Results are given for open universes and fluctuations corresponding to
CDM models with either Harrison-Zeldovich (HZ) or Lyth-Stewart-Ratra-Peebles 
(LSRP) primordial 
energy--density fluctuation power spectrum. Our perturbation theory 
approach provides a distinctive multipole contribution as compared to the 
primary one, the amplitude of the effect being 
very dependent on normalization. For low--$\Omega$ models, the contribution
of the secondary multipoles 
to the radiation power spectrum is negligible both for
standard recombination and reionized scenarios, with the 2--year
COBE--DMR normalization. For a flat universe this contribution is 
$\approx 0.1-10\%$ depending on the reionization history of the universe and
on the normalization of the power spectrum.

\bs
\bs
\noi {\it Subject headings: cosmology: cosmic microwave background - $\Omega<1$
- 2nd order perturbations }
\np
\baselineskip=12pt
 
\noi{\it I. INTRODUCTION}
\medskip

In the absence of a cosmological term ($\Lambda =0$) and assuming standard
recombination, anisotropies in the temperature of the Cosmic Microwave 
Background (CMB) on large angular scales ($\geq (2\Omega ^{1/2})^{\circ }$) 
are generated via the Sachs-Wolfe effect ($\Omega =1$), (Sachs and Wolfe 1967),
or the generalized Sachs-Wolfe effect ($\Omega < 1$). This last case includes 
not only gravitational fluctuations at recombination but the integrated effect 
that depends on curvature of the time-varying potential along the photon 
trajectory (Anile and Motta 1976, Abbott and Schaefer 1986, Traschen and 
Eardley 1986, Gouda et al.1991).

	For smaller angular scales, there are primary contributions coming from
recombination: photon fluctuations and the Doppler effect from last
scattering. However, secondary anisotropies can also arise in the 
microwave sky due to the following physical processes: i) differential 
gravitational redshifts and blueshifts of the growing non-linear density 
fluctuations act on photons travelling to the observer that contribute an 
integrated gravitational effect (Mart\'\i nez-Gonz\'alez, Sanz and Silk, 1990,
1992, 1994), $^\dagger$\footnote{}
{$^\dagger$It is helpful to distinguish between the integrated Sachs-Wolfe 
effect (Sachs and Wolfe 1967; Hu and Sugiyama 1995) that occurs across linear 
fluctuations only 
when $\Omega\neq 1$ from the differential effect across clusters (Rees and 
Sciama 1968) that is present even if $\Omega=1.$ }    
 ii) hot gas inhomogeneously distributed between recombination and 
the present epoch generates anisotropy associated with non-linear flows 
(Vishniac, 1987), and iii) dust present in the early universe, produced in the
process of formation of bound structures, also generates anisotropies 
especially
in the Wien region (Bond et al., 1991). Such effects are important
if the last scattering surface is modified by reionization. An interesting  
possibility is that the secondary temperature fluctuations imprint signatures
that 
differ in angular scale from the primary fluctuations (see section III).

 We will focus in this paper on the first of the above mentioned effects: the
generation of anisotropies by 2nd order density perturbations in an open model.
For the fluctuations, we will considerer a CDM model with null/negligible
baryon content. Recent papers on this topic deal with the hypothesis of a flat
background: 2nd order perturbations were considered by Mart\'\i nez-Gonz\'alez,
Sanz and Silk (1992) whereas some (analytic) models for the non-linear
evolution were treated by Mart\'\i nez-Gonz\'alez, Sanz and Silk (1994).
Numerical simulations dealing with full non-linear evolution for (more 
realistic) 
models such as CDM have been performed by Tuluie and Laguna (1995).  
Seljak (1995) has also done numerical calculations based on N-body simulations
performed by Gelb and Bertschinger (1994) for a flat universe.
The generic conclusion one can extract from this work is that, 
at least for $\Omega = 1$, the level of anisotropy in the CMB generated at 
low-z by quasi-linear and non-linear evolution of the matter is  
$\Delta T/T\simlt 10^{-6}$, except for some unrealistic models where 
growth commences
at very high z.

	We have computed the resulting fluctuations, which are of second order 
in perturbation theory relative to the uniform open cosmological background, 
for the case in which the density fluctuations are Gaussian. We expect 
non--linear contributions to further increase these minimal fluctuations, but
such effects are highly model--dependent. Our perturbation theory approach 
provides a relatively robust prediction: for CDM open models, $\delta T/ T$ 
has a distinctive multipole contribution compared to the one expected in 
standard inflationary models at the linear level. The amplitude of the effect 
is very sensitive to normalization.              
                                     
\bigskip
 
\noi{\it II. THE SECOND ORDER GRAVITATIONAL EFFECT }
\medskip

\noi{\it a) The integrated gravitational effect:}
\smallskip

 We obtained (Mart\'\i nez-Gonz\'alez, Sanz and Silk, 1990) an expression for 
the secondary anisotropies generated by the linear gravitational potential
$\varphi (t,\vec x)$ associated with non-linear density fluctuations $\Delta (t,
\vec x)$. For a flat or open universe with vanishing pressure,

$$\bigl(\Delta T/T\bigr)_{secondary}=2\int^{t_o}_{t_r} dt {\partial\varphi
\over \partial t}(t,\vec x)\ \ \ ,
\ \ \nabla^2 \varphi=6\Omega a^{-1}\Delta(t,\vec x)\ \ \ ,\eqno(1)$$

\noindent where the scale factor $a(t)$ is normalized at the present time 
($a_0=1$) and 
we choose units such that $c=8\pi G=1$ and the Hubble length at the present 
time $d_o =2H_o^{-1} =1$. The previous equation giving the integrated
gravitational effect is valid in the open case only for scales below
the {\it curvature scale}. The line integral must be performed 
along the geodesic associated with the flat or open Friedmann background

$$\vec x(t,\vec n)=\lambda (a)\vec n\ \ \ , 
a(\lambda )\equiv {(1-\lambda)^2\over 1-(1-\Omega)\lambda ^2} \ \ \ ;
\eqno(2)$$

\noindent where $\vec n$ is the unit vector in the direction of observation. 
This integrated effect, except for the factor 2 that comes from general 
relativity, can be understood in terms of Newtonian mechanics: it represents 
the work performed by the photons, 
propagating from recombination to the present time, against the non-static 
gravitational potential $\varphi$.

\noi{\it b) Calculations:}
\smallskip

	Let us consider either an Einstein-de Sitter or open universe as 
background. Perturbation theory up to 2nd order, for vanishing pressure, 
gives 
the following expression for the density fluctuations:

$$\Delta (t,\vec x)=D\delta +D^2\bigl[{5\over 7}\delta^2+\vec\nabla
\delta\cdot\vec\nabla\xi+{2\over 7}\xi_{,ij}\xi^{,ij}\bigr]\ \ \ ,\ \ \
\nabla^2\xi=\delta \eqno(3)$$

\noindent where $\delta (\vec x)\equiv \delta_r (1+z_r)$. The previous 
expression for
$\Delta $ is exact up to second order for a flat model (Peebles, 1980),
whereas it is a good approximation for open models with $\Omega\simgt 0.1$ 
(Bouchet et al. 1993; Catelan et al. 1995). 

Consequently, the second order effect, as given by equation (1), amounts to

$$\bigl(\Delta T/T\bigr)_{secondary}=12\Omega \int^{a_o}_{a_r} da\,\phi(\vec 
x(a))({D\over a})^2(2f-1)\ \ \ ,$$
where
$$\nabla^2\phi(\vec x)={5\over 7}\delta^2+\vec\nabla\delta\cdot\vec\nabla\xi
+{2\over 7}\xi_{,ij}\xi^{,ij}\ \ \ .\eqno(4)$$    

\noindent In the previous equation, $D(a)$ is the growing mode of the 
perturbations 
normalized to $1$ at the present time. For a flat universe: $D = a$, whereas
for an open universe (Peebles, 1980): 

$$D = {g(x)\over g(x_o)}\ \ \ , x = ({1\over \Omega }-1)a\ \ \ , 
f(x) = {dlnD\over dlna}\ \ \ ,$$

$$g(x) = 1+({3\over x})[1+(1+{1\over x})^{1/2}ln((1+x)^{1/2}-x^{1/2})] 
\ \ \ .\eqno(5)$$    
 
The basic function to be calculated for any
experimental set-up is the multipole component $C_l = <a^2_{lm}>$ 
(where temperature fluctuations expanded in spherical harmonics $Y_{lm}$
are given by $\Delta T(\vec n)=\sum_l\sum_{m=-l}^{l}a_{lm}Y_{lm}(\vec n)$), 
given
by the following expression for the second order gravitational effect (we shall
not consider the monopole and dipole in the calculations of the temperature 
anisotropy below):
  
$$C_l={1152\over \pi}{\Omega }^2\int dk k^{-2} P_{(2)}(k)R_l^2(k)\ ,
R_l(k)\equiv \int_0^{\lambda_r} d\lambda {1-(1-\Omega)\lambda \over 
(1-\lambda)^2} D^2(2f-1)j_l(kp)\ ,\eqno(6)$$

\noindent Here, $p\equiv \lambda/[1-(1-\Omega)\lambda^2]$, $j_l$ is the Bessel 
function of fractional order and
$\lambda_r$ is the distance from the observer to the recombination surface.  
The function $P_{(2)}(k)$ is the power spectrum associated with the 
2nd order density perturbation $\delta_2\equiv {5\over 7}\delta^2+
\vec\nabla\delta\cdot\vec\nabla\xi+{2\over 7}\xi_{,ij}\xi^{,ij}$
and is related to the power spectrum $P_\xi $ of the time derivative of the 
potential $\xi$ by $P_\xi={1\over k^4}P_{(2)}(k)$. The second order
perturbation power spectrum $P_{(2)}(k)$ is given in terms of the first order
power spectrum $P(k)$ by the equation (Goroff et al. 1986, Suto and Sasaki,
1991)

$$P_{(2)}(k)={k^3\over 98(2\pi)^2}\int_0^\infty dr P(kr)\int_{-1}^1 dx
P\bigl(k(r^2+1-2rx)^{1/2}\bigr)\Bigl({3r+7x-10rx^2\over r^2+1-2rx}\Bigr)^2\ \ .
\eqno(7)$$

\noindent In the limit of small $k$, there is a cancellation of the three 
terms 
contributing
to $\delta_2$, implying that $P_{(2)}(k)$ has very little power on large 
scales. In particular, the generic behaviour is $P_{(2)}(k)\propto k^4$ for 
small $k$, independently of the primordial power spectrum.  

With regard to  a possible secondary contribution coming from 3rd order
density perturbations for open models (in the case of flat models this does not
exist because the 1st order gravitational potential is static), we 
have the 
following comment: the coupling of the 1st order
gravitational potential $\varphi^{(1)}\propto {D\over a}$ with the 3rd order 
potential $\varphi^{(3)} \propto {D^3\over a}$ gives a kernel for the 
integrated
gravitational effect proportional to ${D^3\over a^2}(f-1)(3f-1)$, whereas 
the coupling of the 2nd order gravitational potential $\varphi^{(2)} \propto 
{D^2\over a}$ with itself gives a kernel for the integrated
gravitational effect proportional to ${D^3\over a^2}(2f-1)^2$. This second
function is always greater than the first one. Moreover, for quasi-flat models
we expect a negligible contribution from the coupling of 1st-3rd
order perturbations because $f\approx 1$, whereas for low-$\Omega$ models the
integrated gravitational effect due to 2nd-2nd order perturbations is produced
at high-z ($75\%$ of the final effect is produced in the interval $[10,10^3]$
for $\Omega = 0.1$, see next section) where $f\approx 1$, and so there is no
practically 1st-3rd order contribution. At smaller z some contribution due to
the 1st-3rd coupling is produced but it is 
estimated to be always bounded by that
due to the 2nd-2nd coupling at low-z, and this is a small fraction of the
final contribution.

\bigskip
	
\noi{\it III. RESULTS AND CONCLUSIONS}
\medskip

We have applied the formalism of the previous section to calculate the 
predicted
amplitudes of the multipole components $C_l = <a^2_{lm}> $. We assume an open
or flat model as background and matter density perturbations corresponding to 
CDM models with either a primordial HZ spectrum $P(k)=Ak$
or a LSRP one (Lyth and Stewart 1990, Ratra and Peebles 1994) and
$\Omega_b=0.05$, h=0.5. Primary anisotropies are normalized to the 2--year
COBE-DMR map as given by the analysis of Cay\'on et al. (1995) for HZ and 
Gorski et al. (1995) for LSRP. However, secondary anisotropies are normalized
to $\sigma_{16}=1$ as this effect is generated by the small scale structure,
$\simlt 100 Mpc$, where maybe the $\sigma_{16}$ normalization is more 
appropriated ($\sigma_{16}$ is the rms density fluctuation at 16 Mpc).
The results of our calculations are presented in the following figures.

We plot in Figure 1 the 2nd order power spectrum $P_2(k)$, calculated
according to eq.(7), for different values of $\Omega$ and normalizing to
$\sigma_{16}=1$. The amplitude of the maximum decreases 
for low $\Omega $ values, and appears at lower $k$.
In Figure 2, we display $C_l$ up to $l =1000$ for different values of 
$\Omega$ due
to 2nd order anisotropies (lower curves) and due to primary anisotropies (upper
curves). The solid, dashed and dotted lines
represent the cases $\Omega = 1, 0.3, 0.1$, respectively. For the 
dashed and dotted curves the results for the HZ spectrum are above the LSRP 
ones 
at $l=100$ in all cases.
With regard to the 2nd order anisotropies, the 
maximum is at $l\approx 250 $ at the level of $10^{-13}$ for all $\Omega$ 
values ($\sigma_{16}=1$).
In the case of COBE-DMR normalization, the level of anisotropy increases by a
factor 5 for $\Omega=1$ (upper solid line) and decreases by 2 and 5 orders of
magnitude for $\Omega=0.3$ and 0.1, respectively.

	We wish to emphasize that for $\Omega = 1$ more than $90\%$ of the 
anisotropy produced by the second order gravitational effect is generated at a 
redshift $z\simlt 10$ (Mart\'\i nez-Gonz\'alez et al. 1992). The case of 
$\Omega < 1$ is similar, for instance if 
$\Omega = 0.1$ about $80\%$ of the effect is generated at a redshift $z<30$ 
(see Figure 3). 
Since reionization could not have produced a substantial 
effect at such low redshifts, for any baryon density consistent with 
primordial nucleosynthesis, our results provide the $minimal$ fluctuations in 
the CMB, independently of the ionization history of the universe.
Notice that for $\Omega=0.1$ there appears a knee in the
generation of the effect with redshift which is due to the change in the
expansion of the universe from being matter dominated to curvature dominated.
The low order multipoles are generated during the curvature dominated phase 
whereas the high order ones are generated during the matter dominated one, as
can be seen in figure (3) for $l=250$. 
These two contributions at relatively high and low redshift are similar to 
the ones appearing at the linear level, the so called SW and ISW effects.

We remark that there is no problem with the influence of non-linear scales in
the second--order calculation because for any
$\Omega $ the maximum of the function $R_l(k)$ (see eq.(6)), is at $k \approx
l\Omega$ and the scales that are contributing to the multipole $l$ have 
$k \simlt l$,
so for the calculated multipoles $C_l$ up to $l = 1000$ only scales $\geq
6h^{-1}\,\rm Mpc$ generate a 2nd order gravitational effect. Hence
no contributions from density fluctuations with density contrast larger
than unity enter in the perturbative calculation.   

The principal conclusion of this analysis is that a deformation ({\it i.e.} a 
peak) at $l \approx
250$ in the 2D radiation power spectrum, $C_l$, would be a signature of the 
2nd order anisotropy  contribution. Detection of such a feature  would allow 
an estimate to be made of the global parameter $\Omega$,
because the first acoustic peak emerging from the primary anisotropy is at 
$l(\Omega)
\simgt 220$. However, such a deformation is negligible for low--$\Omega$ 
models and it is at least two orders of magnitude
below the corresponding primary anisotropy in the flat case. Only a 
possible experiment allowing the $C_l$'s to be determined with an accuracy  
better than $1\%$ could detect such secondary anisotropy.

On the other hand, early reionization of the universe would erase the
primary anisotropy whereas the secondary one could survive with only minor
changes. Figure (4) shows the primary radiation spectrum for several 
reionization histories ranging from standard recombination to 
the case of no recombination  as compared  to the secondary contribution for
$\Omega=1$ and HZ primordial spectrum. The secondary effect amounts to a 
contribution of  $\sim 1-10\%$ in the best case.

	Finally, we remark that the second order gravitational effect is very 
sensitive to normalization because $\Delta T/T$ is proportional to the  
amplitude $A$ of the primordial spectrum.

\medskip

It is a pleasure to thank Wayne Hu for interesting discussions. 
The research of J.S. is supported in part by  grants from  NASA, DOE  and NSF.
E.M.-G. and J.L.S.
are supported by the Spanish DGICYT project PB92-0434-C02-02. L.C. is supported
by a Fulbright fellowship.
E.M.-G. 
thanks the NSF Center for Particle Astrophysics in Berkeley for its 
hospitality and facilities during his stay at Berkeley.
 
\medskip
\np
 
\noi{\it REFERENCES}
                                                            
Abbott, L. F. and Schaefer, R. K. 1986, Ap. J., 308, 546.
 
Anile, A. M. and Motta, S. 1976, Ap. J., 207, 685. 

Bond, J. R., Carr, B. and Hogan, C. J. 1991, Ap. J., 367, 420.
                                             
Bouchet, B., Juszkievicz, R., Colombi, S. and Pellat, R. 1993, preprint.

Catelan, P., Lucchin, F., Matarrese, S. and Moscardini, L. 1995, preprint.

Cay\'on, L., Mart\'\i nez-Gonz\'alez, E., Sanz, J. L., Sugiyama, N. and 
	Torres, S. 1995, MNRAS, to be published.

Gelb, J. and Bertschinger, E. 1994, Ap. J., 436, 467.

Gorski, K. M., Ratra, B., Sugiyama, N. and Banday, A. J., 1995, Ap. J., 
         444, L65.
                                             
Gouda, N., Sugiyama, N. and Sasaki, N. 1991, Prog. Theor. Phys., 85, 1023.

Hu, W. and Sugiyama, N. 1995, Phys. Rev. D., 50, 627.

Goroff, . H.,   Grinstein,  B.,  Rey, S.-J. and  Wise, M. B.
1986, Ap. J., 311, 6.
 
Lyth, D. H. and Stewart, E. D. 1990, Phys. Lett. B, 252, 336.

Mart\'\i nez-Gonz\'alez, E., Sanz, J. L. and Silk, J. 1990, Ap. J., 
	355, L5.

Mart\'\i nez-Gonz\'alez, E., Sanz, J. L. and Silk, J. 1992, Phys. Rev.D, 
	46, 4193.

Mart\'\i nez-Gonz\'alez, E., Sanz, J. L. and Silk, J. 1994, Ap. J., 
	436, 1.

Peebles, P. J. E. 1980, The Large Scale Structure of the Universe, 
	(Princeton, Princeton University Press).

Ratra, B. and Peebles, P. J. E., 1994, Ap. J., 432, L5.
                                             
Rees, M. J. and Sciama, D. W. 1968, Nature, 217, 511.

Seljak, U.,1995, preprint.

Sachs, R. K. and Wolfe, A. N. 1967, Ap.J., 147, 73.

Scott, D. and White, M. 1994, in CMB Workshop: Two years after COBE,
	ed. L. Krauss, World Sci., Singapore, p.214. 

Suto, Y. and Sasaki, M. 1991, Phys. Rev. Lett., 66, 265.

Tuluie, R. and Laguna, P. 1995, Ap.J., 445, L73.

Traschen, J. and Eardley, D. M. 1986, Phys. Rev., 34, 1665. 

Vishniac, E. T. 1987, Ap. J., 322, 597.

\vfill\eject
\noi{\it FIGURE CAPTIONS}
\medskip
 
\item {Figure 1.}  The second--order power spectrum  for the energy--density
perturbations $P_2(k)$ for $\Omega=1,0.3,0.1$ is shown by the solid, dashed and
dotted lines, respectively. The primordial spectrum is normalized to
$\sigma_{16}=1$. 
 
\item {Figure 2.} Radiation Power spectrum  for the primary (upper curves) and 
secondary (lower curves) contributions. $\Omega=1,0.3,0.1$ correspond to solid,
dashed and dotted lines, respectively. The results for the HZ spectrum are
above the LSRP ones at $l=100$ in all cases. The secondary contribution is
normalized to $\sigma_{16}=1$, except for the solid upper curve which
represents the flat case with the COBE-DMR normalization.

\item {Figure 3.} Generation of the multipole $l=250$, with maximum amplitude,
for the case $\Omega=1, 0.3, 0.1$ as a function of redshift (solid, dashed and
dotted lines respectively). 

\item {Figure 4.} Radiation Power spectrum  for several reionization models
with $\Omega=1$ and the HZ spectrum.
Dashed lines from top to bottom correspond to the primary contribution in the
following cases: standard recombination, $\tau=0.5,1$ and no--recombination.
The solid lines represent the secondary contribution which is insensitive to
the recombination history, upper curve is for COBE-DMR normalization and the
lower one for $\sigma_{16}=1$ as in figure 2.
 
\end